\documentclass[10pt,letterpaper]{article}
\usepackage{opex3}

\usepackage{amsmath}
\usepackage{color}
\usepackage{graphicx}
\usepackage{citesort}

\newcommand{\comment}[1]{}
\newcommand{\etal}{et~al.}
\newcommand{\ie}{\mbox{i.\!e.~}}
\newcommand{\eg}{\mbox{e.\!g.~}}
\newcommand{\eqnref}[1]{Eq.~(\ref{#1})}
\newcommand{\tabref}[1]{Table~\ref{#1}}
\newcommand{\figref}[1]{Fig.~\ref{#1}}
\newcommand{\secref}[1]{Section~\ref{#1}}

\newcommand{\eps}{\varepsilon}

\newcommand{\Power}{{\mathcal{P}}}

\newcommand{\total}{{\rm{d}}}
\newcommand{\imag}{{\rm{i}}}

\newcommand{\Group}{\mathcal{G}}
\newcommand{\Subspace}{\mathcal{S}}
\newcommand{\groupelem}{\hat{R}}
\newcommand{\rotation}{\mathcal{R}}
\newcommand{\darst}[1]{{\mathcal{D}^{(#1)}}}
\newcommand{\Span}{{\rm{span}}}
\newcommand{\inline}[1]{\mbox{${#1}$}}

\renewcommand{\vec}[1]{{\bf{#1}}}
\newcommand{\tensor}[1]{{\underline{\bf{#1}}}}

\newcommand{\Cv}{\mathcal{C}_{v}}
\newcommand{\Cvv}{\mathcal{C}_{2v}}
\newcommand{\Cvvv}{\mathcal{C}_{3v}}
\newcommand{\Csixv}{\mathcal{C}_{6v}}
\newcommand{\Cnv}{\mathcal{C}_{nv}}
\newcommand{\boundary}{\mathcal{B}}

\begin{document}

\title{Formal selection rules for Brillouin scattering in integrated waveguides and structured fibers}
\author{C. Wolff,$^{1,2,\ast}$ M.~J.~Steel,$^{1,3}$ and C.~G.~Poulton$^{1,2}$}
\address{
  \centering
  $^1$ Centre for Ultrahigh Bandwidth Devices for Optical Systems (CUDOS), Australia
  \newline
  $^2$ School of Mathematical Sciences, University of Technology Sydney, NSW 2007, Australia
  \newline
  $^3$ Department of Physics and Astronomy, Macquarie University Sydney, NSW 2109, Australia
}%

\email{$^{\ast}$christian.wolff@uts.edu.au}

\date{\today}

\begin{abstract}
We derive formal selection rules for Stimulated Brillouin Scattering (SBS) 
in structured waveguides.
Using a group-theoretical approach, we show how the waveguide symmetry 
determines which optical and acoustic modes interact for both forward and 
backward SBS. 
We present a general framework for determining this interaction and give 
important examples for SBS in waveguides with rectangular, triangular and 
hexagonal symmetry. 
The important role played by degeneracy of the optical modes is illustrated. 
These selection rules are important for SBS-based device design and for a 
full understanding the physics of SBS in structured waveguides. 
\end{abstract}


%%%%%%%%%%%%%%%%%%%%%%%%%%%%%%%%%%%%%%%%%%%%%%%%%%%%%%%%%%%%%%%%%%%%%%%%%%%%%%%%
%%%%%%%%%%%%%%%%%%%%%%%%%%%%%%%%%%%%%%%%%%%%%%%%%%%%%%%%%%%%%%%%%%%%%%%%%%%%%%%%
\section{Introduction}

Stimulated Brillouin Scattering (SBS) is the coherent nonlinear interaction
between light and sound in matter~\cite{Boyd2003}.
It was initially predicted by Brillouin~\cite{Brillouin1922} and subsequently
observed in quartz shortly after the invention of the laser~\cite{Chiao1964}.
It is a third-order nonlinear process with several technological applications.
Under suitable conditions including narrow pump sources, the effect
can be very strong---it has the lowest threshold of all nonlinear
processes in standard optical fibres~\cite{Agrawal}.
Still, long interaction lengths are usually required to obtain appreciable 
interaction at moderate light intensities. 
As a result, most existing or proposed applications such as Brillouin 
lasers~\cite{Kabakova2013,Hu2014}, all-optical isolators~\cite{Huang2011} or in
microwave photonics~\cite{Morrison2014} are based on optical fibers or on-chip
waveguides.
The interaction in conventional fibers is dominated by electrostriction and 
the photoelastic effect, which, until relatively recently, were viewed as the 
only significant processes underlying SBS.

Due to the nonlinear nature of the process, however, the interaction can be 
dramatically increased in strength by confining the optical mode more tightly, 
thereby increasing the field intensity for the same transmitted power.
Consequently, theoretical and experimental investigations of SBS in
structured fibers and integrated photonic waveguides have recently received 
considerable 
attention~\cite{Dainese2006,Kang2009,Florea2006,Shin2013,Rakich2010,Rakich2012,Pant2011}
with new physical mechanisms~\cite{Rakich2010,Rakich2012,Wolff2014} being 
found to contribute to the SBS gain.
In addition to the intensity enhancement, such structured waveguides further
differ from traditional systems such as step-index fibers in that they support
a broad variety of acoustic and optical modes. 
As solutions to linear wave equations, both sets of modes must respect the
waveguide's symmetries (rectangular for many integrated waveguides,
triangular or hexagonal for many photonic crystal or suspended core fibers), 
as expressed by their corresponding point groups.
These symmetries play a key role in the question of whether a given combination 
of optical and acoustic modes can have non-zero SBS-interaction.
This fact has been recently studied for the specific case of a
rectangular waveguide; Qiu~\etal~\cite{Qiu2012} numerically computed SBS gains 
for a range of possible interactions and found starkly different gains for 
different mode combinations.

In this paper, we approach SBS interactions on a formal level using group
representation theory. 
We derive clear selection rules for SBS in rectangular, triangular and 
hexagonal waveguides, these being the most common point-groups arising in 
photonic waveguides.
Although the formalism of group theory is a well-established approach in 
studying physical interactions of all types, its use in multi-wave interactions 
in the context of photonic waveguides has been relatively limited.
The definitive categorization of symmetry classes for optical modes was given 
by McIsaac~\cite{McIsaac1975a,McIsaac1975b}, however the acoustic modes 
have not been similarly analyzed, and a systematic tabulation of the different 
types of acoustic modes, including their symmetries and possible degeneracies,
is one aim of the present study. 
More importantly, SBS involves interactions between optical and acoustic modes
in nonlinear combination, and the correct conjunction of modes that will 
permit a non-zero contribution to the gain is a non-trivial problem. 
Here, we systematically show which acoustic modes can be used to drive SBS
transitions between optical modes, and conversely, which optical modes can be 
used to generate desired acoustic vibrations. 

To help motivate our treatment, we note that
  SBS in structured waveguides is an unusually complex nonlinear interaction, because of the complicated
  nature of the acoustic dispersion relation which features many modes of different 
  symmetry at moderate acoustic frequencies. In contrast, for SBS in step index fibers, only one acoustic mode is relevant, while for other nonlinearities, acoustic modes of course do not enter at all.
  In contrast to the optical modes (which can be seeded if necessary), there is
  limited direct control over the type of acoustic mode that is excited due
  to the small acoustic propagation length.
  Instead, the active acoustic mode is selected by the waveguide's geometry and the symmetry 
  properties of the optical modes.
  While brute-force computations of coupling integrals are always an option,
  a refined knowledge of this interplay of the modes' and the waveguide's symmetries
  can be of great value in disentangling the acousto-optic coupling dynamics in 
  complex geometries. It  can also indicate which symmetries (or perhaps near symmetries,)
  of a system are responsible for an unfavorably small or strong coupling.
  As an indication of its potential power, the symmetry analysis of SBS is analogous 
  to the situation in Raman molecular spectroscopy~\cite{hollas2004}, where the the symmetry properties of the molecule under study determine the presence or absence
  of a given resonant line for given input and output polarization.
  Indeed, the relevant overlap integrals (see below) have rather similar forms.
  
Our selection rules give important physical insight into the dominant 
couplings in SBS-active experiments, and can be used to greatly reduce the 
number of mode combinations that have to be considered in gain calculations. 
For example, we show that for backwards SBS, which generally involves 
transitions between the same optical mode, no flexural or torsional acoustic 
modes can possibly contribute to the gain in rectangular or hexagonal 
structures, whereas flexural modes can contribute in triangular waveguides.
Knowledge of the appropriate symmetries of the acoustic modes can also simplify 
the design of SBS-active waveguides, because once the symmetries of the 
interacting modes are known, the computational domains can be restricted by 
imposing appropriate boundary conditions along planes of high 
symmetry~\cite{McIsaac1975a}. 
This of course significantly reduces the calculation time, but more 
importantly it greatly simplifies the isolation of degenerate or 
near-degenerate modes.

The paper is organized as follows:
In Sections 2 we recall some key results about opto-acoustic coupling in SBS 
processes.  
In Section 3, we briefly review the relevant fundamentals of group 
representation theory.
In Section 4, we formulate the acousto-optic interaction in a way that
is convenient for a symmetry analysis, and thereby derive a necessary 
condition on the mode symmetries to permit non-zero SBS-gain.
Up to this point, the paper applies to any waveguide symmetry, including
circular waveguides and more exotic cases such as five-fold symmetry.
In Section 5, we apply this to the aforementioned three important symmetry
classes (rectangular, triangular, hexagonal) and provide examples for
combinations of modes with non-vanishing interaction.
Section 6 concludes our paper.

\begin{figure}
  \centering
  \includegraphics[width=0.92\columnwidth]{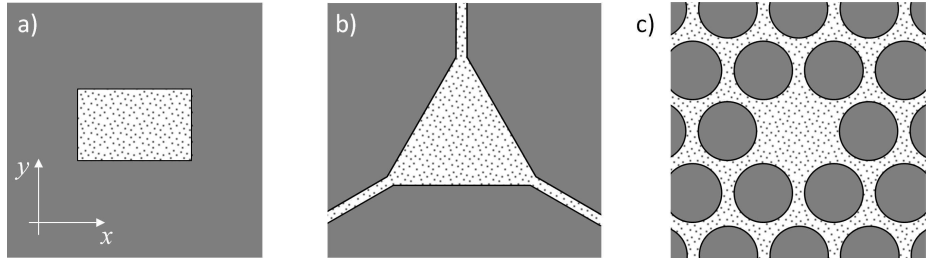}
  \caption{
    Illustration of three different waveguide groups, divided according 
    to the waveguide symmetry; 
    a) rectangular waveguides ($\Cvv$ group),
    b) triangular waveguides ($\Cvvv$ group) and 
    c) hexagonal waveguides ($\Csixv$ group).
  }
  \label{fig:geometries}
\end{figure}

%%%%%%%%%%%%%%%%%%%%%%%%%%%%%%%%%%%%%%%%%%%%%%%%%%%%%%%%%%%%%%%%%%%%%%%%%%%%%%%%
%%%%%%%%%%%%%%%%%%%%%%%%%%%%%%%%%%%%%%%%%%%%%%%%%%%%%%%%%%%%%%%%%%%%%%%%%%%%%%%%

\section{Opto-acoustic coupling in SBS}
We consider longitudinally invariant opto-acoustic waveguides oriented along 
the $z$-axis, and possessing some degree of rotation or reflection symmetry in 
the $x$-$y$-plane (see \figref{fig:geometries}).
The SBS process can be described either in terms of field perturbations or in
terms of optical forces and the mechanical displacement field.
The descriptions are equivalent~\cite{Wolff2014}; here we choose to work with the optical forces.
The coupling between optical and acoustic modes is proportional to the overlap
integral
\begin{align}
\label{eq:Qdef}
  Q = \int \total^2 r \ \vec u^\ast(\vec r) \cdot \vec f(\vec r) = 
  \int \total^2 r \ w(\vec r),
\end{align}
where $\vec u$ is the displacement field pattern of the relevant acoustic mode
and $\vec f$ is the contribution of the optical force field (due to the 
two interacting optical modes) that is phase-matched to the acoustic mode. 
The integration is carried out over the cross section of the waveguide.
Most key properties of the SBS interaction follow from knowledge of $Q$.
For example, in long waveguides at steady state, the conventional SBS power
gain parameter $\Gamma$ is given by 
\begin{equation}
  \Gamma = \frac{2 \omega \Omega |Q|^2}{\Power^3} 
  \Re \left\{ \frac{1}{\alpha - \imag \kappa} \right\},
\end{equation}
where $\kappa$ is a detuning from phase-matching, $\alpha$ is the attenuation 
coefficient of the phonon mode, and the optical and acoustic modes have been 
assumed to carry one unit $\Power$ of power.  
Thus, maximizing the efficiency of calculation and understanding of the overlap 
$Q$ underlies effective manipulation of SBS as a nonlinear tool.

As discussed in a number of recent works~\cite{Wolff2014,Rakich2010,Rakich2012} 
the force field $\vec f$ has two contributions: electrostriction and 
boundary radiation 
pressure, which lead to separate contributions to the coupling density 
\begin{align}
\label{eq:wdef}
  w(\vec r) = &
  \underbrace{
    \eps_0 [\eps_r(\vec r)]^2 \sum_{ijkl} p_{ijkl}(\vec r) [e^{(1)}_i]^\ast 
    e^{(2)}_j \partial_k u^\ast_l 
  }_{\text{electrostriction}}
  \ + \ 
  \underbrace{
    \eps_0 \left(\eps^{(a)}_r - \eps^{(b)}_r\right) \sum_{ijkl} M_{ijkl} 
    [e^{(1)}_i]^\ast e^{(2)}_j n_k u^\ast_l
  }_{\text{radiation pressure}} .
\end{align}
In the first term, which describes electrostriction, $\epsilon_0$ is the 
permittivity of free space, $\epsilon_r(\vec r)$ and $\tensor p(\vec r)$
are the local relative dielectric constant and fourth rank photoelastic 
tensor, and $\vec e^{(1,2)}$ are the complex mode amplitudes of the two 
optical fields.
The second term, describing radiation pressure, is expressed in terms
of field value limits on the interior waveguide boundary through
the waveguide local normal vector $\vec n$,
the dielectric constants $\epsilon_r^{(a,b)}$ of the core ($a$) and 
cladding ($b$) media and the fully invariant 
tensor~\cite{Wolff2014,Rakich2010}
\begin{align}
\label{eq:Mdef}
  M_{ijkl}(\vec r) = & 
  \left[(1 - \delta_{ik}) (1 - \delta_{jk}) - \frac{\eps^{(a)}_r}{\eps^{(b)}_r} 
  \delta_{ik} \delta_{jk} \right] \delta_{kl} \ \delta(\vec \in \boundary)
  = M_{jikl} = M_{klij}  .
\end{align}
Here $\delta_{ij}$ is the Kronecker symbol and $\delta(\vec r \in \boundary)$ 
is a slightly simplified notation for the Dirac distribution of a function
that has its zeros on the boundaries $\boundary$ between different dielectrics.
This distribution $\delta(\vec r \in \boundary)$ expresses the fact that 
radiation pressure is confined to the material interfaces.
More technically, it is required in order to formulate the more naturally
arising contour integral in the form of the area integral in 
\eqnref{eq:wdef}.
A third term is caused by the interaction between the magnetic optical fields
and a dynamic magnetization caused by the motion of the material~\cite{Wolff2014}.
This term can be effectively treated as an antisymmetric part of the 
photoelastic tensor $\tensor p$~\cite{Nelson1971} and is typically weak.

The primary goal of this paper is to use symmetry arguments to 
provide a systematic and efficient procedure
for determining whether the integral in Eq.~\eqref{eq:Qdef}
is non-vanishing or not.

%%%%%%%%%%%%%%%%%%%%%%%%%%%%%%%%%%%%%%%%%%%%%%%%%%%%%%%%%%%%%%%%%%%%%%%%%%%%%%%%
%%%%%%%%%%%%%%%%%%%%%%%%%%%%%%%%%%%%%%%%%%%%%%%%%%%%%%%%%%%%%%%%%%%%%%%%%%%%%%%%
\section{Symmetry properties of optical and acoustic modes}
\label{sec:symmprop}

To proceed, we must recall a number of properties of the representation
theory of discrete point groups (see for example \cite{Lax1974,Dresselhaus2008}). 
Formally, let $\groupelem$ be a point symmetry operator, \ie a rotation or 
mirror operation about the (longitudinal) $z$-axis, 
and let $\rotation$ be the corresponding 
$3\times3$-matrix that performs this operation in Cartesian space.
The image of a scalar function $\phi(\vec r)$ under $\groupelem$ is found 
by transforming the underlying coordinate system $\vec r = (x,y,z)$:
\begin{align}
  \groupelem \phi(\vec r) = \phi(\rotation^{-1} \vec r).
\end{align}
If the function is vector or tensor-valued, these values must also be transformed.
For a first rank tensor (\ie vector) field $\vec v(\vec r)$ (such as the 
electric field or the mechanical displacement field) and a second rank tensor 
field $\tensor T(\vec r)$ (\eg the stress tensor) the transformation becomes:
\begin{align}
  [\groupelem \vec v(\vec r)]_i = 
  \sum_j v_j(\rotation^{-1} \vec r) \rotation_{ji} \ ; \quad \quad
  [\groupelem \, \tensor T(\vec r)]_{ij} = 
  \sum_{kl} T_{kl}(\rotation^{-1} \vec r) \rotation_{ki} \rotation_{lj} \ .
\end{align}
Pseudotensorial quantities such as the magnetic field differ from this transformation by an 
additional factor $\det \rotation = \pm1$.

\begin{table*}[t]
  \caption{
    Relevant parts of the character tables for the groups 
    $\Cvv$, $\Cvvv$ and $\Csixv$.
    These tables are standard in many textbooks on group 
    theory and we reproduce them here for convenience following 
    the notation of~\cite{Lax1974}.
  }
  \label{tab:representations}

  \vspace{2mm}
  \noindent
  \begin{minipage}[t][][b]{0.30\textwidth}
    \resizebox{\columnwidth}{!}{
      \begin{tabular}{c|rrrr}
        $\Cvv$ & $E$ & $C_{2}$ & $\sigma_y$ & $\sigma_x$ \\
        \hline
        $A  $ & $ 1$ & $ 1$ & $ 1$ & $ 1$ \\
        $B_1$ & $ 1$ & $-1$ & $ 1$ & $-1$ \\
        $B_2$ & $ 1$ & $ 1$ & $-1$ & $-1$ \\
        $B_3$ & $ 1$ & $-1$ & $-1$ & $ 1$ \\
      \end{tabular}
    }
  \end{minipage}
  \hspace{0.01\textwidth}
  \begin{minipage}[t][][b]{0.24\textwidth}
    \resizebox{\columnwidth}{!}{
      \begin{tabular}{c|rrr}
        $\Cvvv$ & $E$ & $C_{3}$ & $\sigma_x$ \\
       \hline
       $A_1$ & $ 1$ & $ 1$ & $ 1$ \\
       $A_2$ & $ 1$ & $ 1$ & $-1$ \\
       $E$   & $ 2$ & $-1$ & $ 0$ \\
      \end{tabular}
    }  
  \end{minipage}
  \hfill
  \begin{minipage}[t][][b]{0.42\textwidth}
    \resizebox{\columnwidth}{!}{
      \begin{tabular}{c|rrrrrr}
        $\Csixv$ & $E$ & $2C_{3}$ & $3\sigma_y$ & $C_2$ & $2C_6$ & $3\sigma_x$ \\
        \hline
        $A_1$ & $ 1$ & $ 1$ & $ 1$ & $ 1$ & $ 1$ & $ 1$ \\
        $A_2$ & $ 1$ & $ 1$ & $-1$ & $ 1$ & $ 1$ & $-1$ \\
        $B_1$ & $ 1$ & $ 1$ & $ 1$ & $-1$ & $-1$ & $-1$ \\
        $B_2$ & $ 1$ & $ 1$ & $-1$ & $-1$ & $-1$ & $ 1$ \\
        $E_1$ & $ 2$ & $-1$ & $ 0$ & $-2$ & $ 1$ & $ 0$ \\
        $E_2$ & $ 2$ & $-1$ & $ 0$ & $ 2$ & $-1$ & $ 0$ \\
      \end{tabular}
    }
  \end{minipage}
\end{table*}

If the waveguide in all its aspects (specifically including its tensorial material 
parameters,) is invariant with respect to a given set of point symmetry operations, 
these operations form a group $\Group$ with respect to composition.
Common examples are the $\Cvv$, $\Cvvv$ and $\Csixv$ groups that apply to 
waveguides with rectangular, triangular and hexagonal symmetry. 
The elements of the $\Cnv$ groups consist of the identity operation ($E$), 
all multiples of the $2 \pi/n$ rotation ($C_n$), and $2n$ reflection operations 
that are created by successively applying half the fundamental rotation operation, 
\ie by applying the $\pi/n$-rotation to the $x$-mirror operation ($\sigma_x$), 
which maps any vector $(x, y, z)$ to $(-x, y, z)$.
For each excitation frequency $\omega$, the set of (optical or acoustic) modes 
$\{ f^{(i)}(\vec r) \}$ with the same wavenumber forms a linear subspace 
$\Subspace_f$ that is invariant under all operations in $\Group$.  
Each symmetry operation $\groupelem \in \Group$ can therefore be expressed as 
an expansion in the basis of the mode functions $f^{(i)}(\vec r)$:
\begin{align}
  \groupelem f^{(i)} = \sum_j f^{(j)} \darst{\groupelem}_{ji} \ .
\end{align}
In general $\darst{\groupelem}$ are matrices, however in the non-degenerate 
case, in which the set $\{ f^{(i)}(\vec r) \}$ consists of a single function, 
the matrices reduce to scalar phase factors. 
In the degenerate case, the matrices $\darst{\groupelem}$ depend on the choice of
basis functions $f^{(i)}(\vec r)$ and are unitary for any ortho-normal basis.

The set of matrices $\darst{\groupelem}$ for all elements 
$\groupelem \in \Group$ is known as a (matrix) {\em representation} of the 
underlying symmetry group $\Group$, which means that these matrices fulfill the 
multiplication rules of the abstract group elements.
Different mode types induce specific representations that are associated with 
the modal symmetry. 
As the matrices themselves depend on the choice of basis functions, they are 
ill-suited to characterize the symmetry of a degenerate subspace.
Instead, the natural choice are the matrices' traces, which are invariant under 
similarity transformations and called the \emph{characters} of the representation 
matrices.
For a non-degenerate mode that is even with respect to all elements 
$\groupelem \in \Group$, that is, it is unchanged under all operations in the 
group, the representation induced by the mode consists of 
$\darst{\groupelem} = 1$ for each element $\groupelem$. 
This trivial representation, denoted by $A$ or by $A_1$ exists for each group 
and corresponds to monopole-like fields.  
We should stress that the mere existence of this representation does not mean 
that the (optical or acoustic) wave equation will necessarily have a physical 
solution of this symmetry at any given frequency
(indeed the fundamental mode of a dielectric waveguide is only rarely of $A_1$ 
type~\cite{Bassett2002,deSterke1994}).
Representations that are characterized by sign changes under the various 
operations of the group (rotations and reflections) correspond to higher-order 
multipole fields or superpositions thereof and are denoted $A_2$, $B$-type or 
$E$-type representations, the latter denoting degenerate representations.
Table~\ref{tab:representations} lists the characters of all irreducible 
representations of the three point groups under consideration, for sets of 
operations that are sufficient to unambiguously identify the representations 
based on their characters.  
These tables are reproduced from~\cite{Lax1974} (and found in many other texts).
The degenerate representations $E$ and $E_1$ of the groups 
$\Cvvv$ and $\Csixv$, respectively, are particularly important since the 
fundamental optical mode of a dielectric waveguide with threefold or 
sixfold symmetry is of this type in practical situations involving index-guided
waveguides.

The acousto-optic overlap integral involves a product of three modes (two optical 
and one acoustic), so we need the symmetry properties of products.
The product of two modes $f^{(i)}(\vec r) \cdot g^{(j)}(\vec r)$ from two 
invariant subspaces $\Subspace_f$ and $\Subspace_g$ lies inside the direct 
product subspace spanned by every possible product 
$\Subspace_f \otimes \Subspace_g = \Span\{ f^{(i)}(\vec r) g^{(j)}(\vec r) \}$.
If $\Subspace_f$ is one-dimensional, the characters of 
$\Subspace_f \otimes \Subspace_g$ are obtained by simply multiplying the 
characters of $\Subspace_g$ with those of $\Subspace_f$ (and analogously if 
$\Subspace_g$ is one-dimensional).
If both subspaces are degenerate (with dimensions $n$ and $m$), the product 
space is $n\times m$-dimensional and usually reducible into the few irreducible 
representations of the point group, \ie it can be decomposed into a number of 
independent smaller subspaces.
In the context of discrete symmetry groups for waveguides, this is always 
the case.
The details of this decomposition depend on whether 
$\Subspace_f \neq \Subspace_g$ (general case), or whether the functions 
$f^{(i)}(\vec r)$ and $g^{(j)}(\vec r)$ are identical or orthogonal 
to each other within the same degenerate subspace $\Subspace_f = \Subspace_g$.
Within the context of SBS, the latter appears if the two optical modes have
(up to the Stokes shift) the same frequency and wave number.
``Identical'' in this context means that the optical mode is red-shifted in
frequency but its polarization state is not affected.
``Orthogonal'' refers for example to the case that the polarization state is 
completely changed, \ie that the source and destination states are orthogonal.
The general rule applies whenever one of the involved modes has a degenerate 
partner that is not excited by the SBS-process under consideration.
For example, the flexural acoustic modes in $\Cvvv$ or $\Csixv$ systems always
appear in pairs, even though only one of these mode may contribute in a
specific setup.

For the two point groups under consideration that feature degenerate 
representations, the product decompositions are listed in 
\tabref{tab:prod_decomp}.
Again, these tables are reproduced from~\cite{Lax1974}.

%%%%%%%%%%%%%%%%%%%%%%%%%%%%%%%%%%%%%%%%%%%%%%%%%%%%%%%%%%%%%%%%%%%%%%%%%%%%%%%%
\subsection{Examples for optical and acoustic modes}

\begin{figure}
  \centering
  \includegraphics[width=0.77\textwidth]{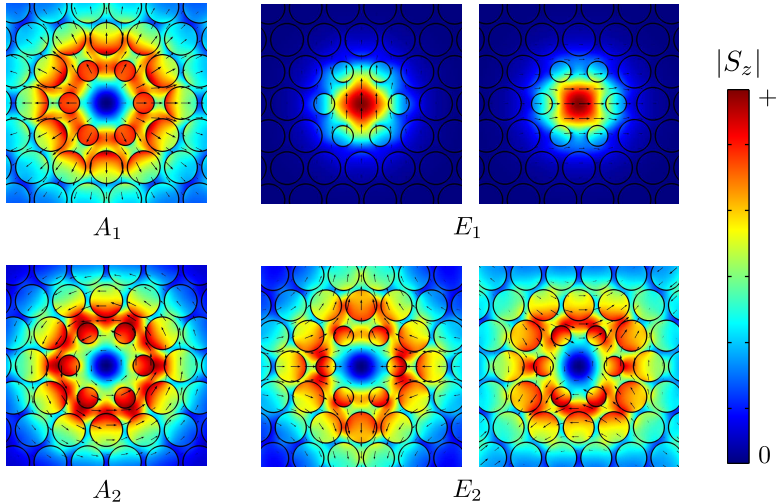}
  \caption{
    Electric field patterns for the fundamental optical modes in a structured 
    silica fiber with hexagonal symmetry.
    The arrows represent the transversal field components. 
    The color represents the time-averaged power density directed along the 
    waveguide axis (\ie the $z$-component of the time-averaged Poynting vector).
    The color scale is in arbitrary units.
  }
  \label{fig:optical_examples}
\end{figure}

The symmetry properties of optical waveguides are well-known \cite{McIsaac1975a}. 
Here we show some specific examples to illustrate how the group-theoretical 
representation relates to the more widely-used categories of waveguide modes. 
In \figref{fig:optical_examples} we show examples of optical modes of a
photonic crystal fiber, with a structure belonging to the $\Csixv$ point group. 
The two fundamental modes have $E_1$-symmetry, and correspond to the 
fundamental linearly $x$- and $y$-polarized modes (often classified as 
HE$_{11}$ modes, by way of analogy with the step-index fiber). 
These modes are the most relevant for optical experiments, since they are 
readily coupled into from free space or other waveguides.
Higher order modes are also possible in this structure; in this specific 
configuration and at this frequency these are all cladding modes, however 
these mode types can become important for large-core structures. 
The first higher-order $x$ and $y$ polarized modes are of $E_2$ type and are 
also degenerate.
The $A_1$ type mode in \figref{fig:optical_examples} has an electric field 
pointing radially outward, this mode possessing even vectorial symmetry both 
under all reflections and rotations, in accordance with 
\tabref{tab:representations}.
The $A_2$-type mode has odd symmetry under all reflections and even symmetry 
under all rotations; this ``donut'' mode rotates its polarization with respect 
to angle and possess a phase singularity at the origin, and is analogous to 
the TE$_{01}$ mode of the step-index fiber.
The $B_1$ and $B_2$ type modes arise rarely in practical situations and so we 
have omitted these in the figure; a sketch of the properties of these modes 
can be found in \figref{fig:C6v}.
In all cases we note that it is the transformation properties of the electric 
field which determine the classification, the magnetic field being a pseudo-vector.

\begin{figure}
  \includegraphics[width=\textwidth]{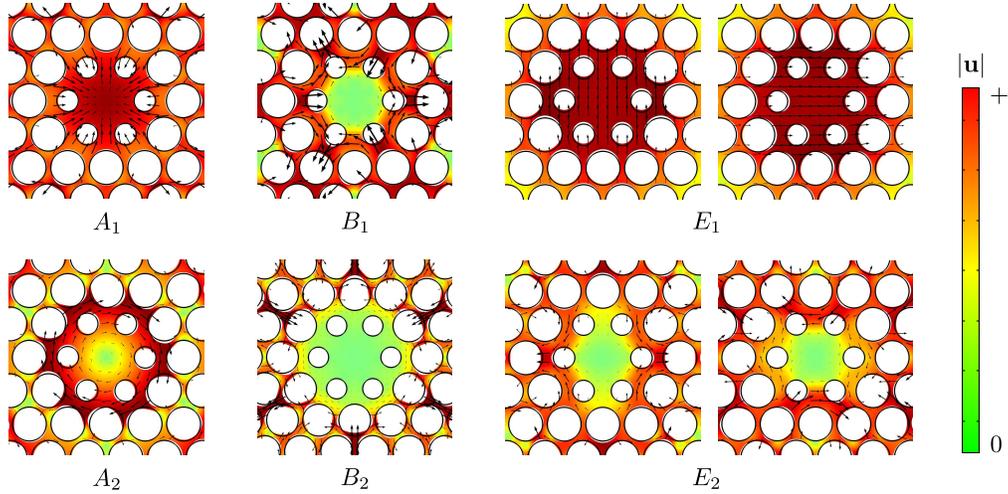}
  \caption{
    Examples for the symmetries of acoustic modes in a structured silica fiber
    with hexagonal symmetry.
    The arrows and the deformation plot indicate the in-plane components
    $\vec u_\perp = (u_x, u_y, 0)$ of the mechanical displacement field.
    The color represents the absolute value of the total displacement field
    $|\vec u| = \sqrt{|u_x|^2 + |u_y|^2 + |u_z|^2}$.
    The color scale is in arbitrary units.
  }
  \label{fig:acoustic_examples}
\end{figure}

The symmetry properties of acoustic modes are perhaps less known within the 
photonics community.
In \figref{fig:acoustic_examples} we show examples of the different types of 
acoustic modes, again for the $\Csixv$ point group. 
The $A_1$ type mode corresponds to a longitudinal-like acoustic mode, because 
the dominant motion is in the waveguide direction.
The $A_2$-type mode is a torsional-like mode, and the degenerate pair of 
$E_1$ modes are the orthogonally-polarized fundamental flexural modes. 
The additional degenerate pair of $E_2$-type acoustic modes either stretches 
or shears the core in the horizontal direction, rather than displacing the 
core as a whole.
This representation covers \eg accordion-like ``breathing modes'', which 
simultaneously stretch the core horizontally while compressing it vertically.
The acoustic $B_1$ and $B_2$ modes describe non-degenerate higher-order 
vibrations. 
The $B_1$ modes displace the material either inwards or outwards along 
symmetry planes separated by 120 degrees, while displacing material in the 
opposite direction along the alternate symmetry planes. 
The $B_2$ modes act in a similar way to the $B_1$ modes, but with the symmetry 
planes rotated by 30 degrees.

%%%%%%%%%%%%%%%%%%%%%%%%%%%%%%%%%%%%%%%%%%%%%%%%%%%%%%%%%%%%%%%%%%%%%%%%%%%%%%%%
%%%%%%%%%%%%%%%%%%%%%%%%%%%%%%%%%%%%%%%%%%%%%%%%%%%%%%%%%%%%%%%%%%%%%%%%%%%%%%%%

\section{Formal selection rule}

\begin{table*}[t]
  \caption{
    Relevant parts of the product 
    decomposition tables for $\Cvvv$ and $\Csixv$.
    The abbreviations ``i.d.m.'' and ``o.d.m.'' 
    in the decomposition tables stand for 
    ``identical degenerate modes'' and 
    ``orthogonal degenerate modes'', respectively.
    See \secref{sec:symmprop} for details.
    Again, these tables are standard in many textbooks on group 
    theory such as~\cite{Lax1974}.
  }
  \label{tab:prod_decomp}

  \vspace{2mm}
  \noindent
  \begin{tabular}{lr}
    Product decomposition for $\Cvvv$ \\
    \hline
    $E \otimes E = A_1 \oplus E $ & (i.d.m.) \\
    $E \otimes E = A_2 $  & (o.d.m.) \\
    $E \otimes E = A_1 \oplus A_2 \oplus E $ & (general) 
    \\
    & \\
    & \\
    & \\
    & 
  \end{tabular}
  \hfill
  \begin{tabular}{lr}
    Product decomposition for $\Csixv$ \\
    \hline
    $E_1 \otimes E_1 = A_1 \oplus E_2 $ & (i.d.m.) \\
    $E_1 \otimes E_1 = A_2 $  & (o.d.m.) \\
    $E_1 \otimes E_1 = A_1 \oplus A_2 \oplus E_2 $ & (general) \\
    $E_2 \otimes E_2 = A_1 \oplus E_2 $ & (i.d.m.) \\
    $E_2 \otimes E_2 = A_2 $  & (o.d.m.) \\
    $E_2 \otimes E_2 = A_1 \oplus A_2 \oplus E_2 $ & (general) \\
    $E_1 \otimes E_2 = B_1 \oplus B_2 \oplus E_1 $ & (general) \\
  \end{tabular}
\end{table*}

The term ``selection rule'' usually refers to a set of constraints on the 
symmetry properties of physical states to allow for interaction.
The best known example of these are the conditions for the atomic states 
between which optical transitions can occur.
For that problem, the point group is the group of continuous rotations and the
irreducible representations are distinguished by the two angular momentum 
quantum numbers.
Transitions can occur if the product of the two electronic representations 
and the representation decomposition of the electric dipole operator has a
trivial (\ie with total angular momentum $L=0$) contribution, which is just
another formulation of the well known optical selection rules 
$\Delta L=\pm1$ and $-1 \leq \Delta m \leq 1$.
Similar rules for discrete symmetry groups of atoms in solids are known from
the context of crystal field theory.

In this section, we formulate the problem of acousto-optic interaction in
a very similar picture, where the optical waveguide modes are analogous to the
electronic eigenstates and the acoustic wave corresponds to the electric 
dipole operator.
Using this, we formulate the opto-acoustic selection rule in its most general 
form in analogy to the problem of optical transitions in atoms. 
To this end, we now analyse the symmetry properties of the overlap integral 
\eqnref{eq:wdef}.
The photoelastic tensor $\tensor p$ is invariant under the symmetry operations, 
because we assumed the waveguide (including all material parameters) to be so.
This is strictly true for cubic materials in rectangular geometries and for 
isotropic materials such as silica or various glasses under all circumstances.
The tensor $\tensor M$ is invariant by construction.
The partial derivative and the normal vector distribution $\vec n$ induce the
trivial group representations.
Thus, we can formally write Eq.~\eqref{eq:wdef} in the form
\begin{align}
  w = \sum_{ijk} N_{ijk} [e^{(1)}_i]^\ast e^{(2)}_j u^\ast_k \ ,
  \label{eqn:integrand}
\end{align}
where the third rank tensor operator $\tensor N$ is invariant under all 
waveguide symmetries, and induces the trivial representation $A$ or $A_1$, 
depending on the precise group.

To formulate the selection rule for SBS, we first write the integrand as the 
superposition of symmetric functions 
\inline{w = \sum_\alpha w_\alpha}, where each index 
$\alpha$ corresponds to one irreducible representation.
The integrals over all symmetric contributions are zero except for the 
trivial representation~\cite{Lax1974}.
That is, the coupling integral $Q$ can be only non-zero for modes whose product 
contains the trivial representation (note however that this does not imply that 
the integral over a trivial representation is always non-zero).
To this end, we first determine the symmetry properties of the product between 
the two optical modes, either by simply multiplying the characters from 
Table~\ref{tab:representations} in the case of at least one non-degenerate mode
or by using Table~\ref{tab:prod_decomp} if both modes are degenerate.
In the same way, we then combine every irreducible representation from this first
product with the acoustic representation.
If this yields a trivial representation, coupling is possible.

To illustrate this evaluation of representation products, we give two examples:
First assume that in a rectangular waveguide the optical representations are
$B_1$ and $B_3$ and the acoustic representation is also $B_3$.
The product of the characters of $B_1$ and $B_3$ result in the characters of
$B_2$, thus the product of the two optical modes has $B_2$-symmetry.
The product of these characters and those of the acoustic $B_3$-representation
indicate that the product of all three modes has $B_1$-symmetry, and hence that
the overlap integral vanishes.

Second, assume that in a hexagonal waveguide the optical representations are
$E_1$ and $E_2$ and the acoustic representation is $B_1$.
From Table~\ref{tab:prod_decomp} we can see that the product of the optical 
modes is a superposition of a $B_1$-, a $B_2$- and a degenerate 
$E_1$-representation.
Each of these representations are now individually combined with the acoustic
$B_1$-representation to find the decomposition of the total integrand:
it is a superposition of an $A_1$-, an $A_2$- and an $E_2$-representation.
The integral over the latter two contributions vanishes, but the integral over
the first term can be non-zero.
Thus coupling is possible but may depend on the exact choice of the optical modes
within their respective $E_1$- and $E_2$-subspaces.

%%%%%%%%%%%%%%%%%%%%%%%%%%%%%%%%%%%%%%%%%%%%%%%%%%%%%%%%%%%%%%%%%%%%%%%%%%%%%%%%
%%%%%%%%%%%%%%%%%%%%%%%%%%%%%%%%%%%%%%%%%%%%%%%%%%%%%%%%%%%%%%%%%%%%%%%%%%%%%%%%
\section{Examples for SBS-active mode combinations}

We now examine the three most important cases in more detail.
 
%%%%%%%%%%%%%%%%%%%%%%%%%%%%%%%%%%%%%%%%%%%%%%%%%%%%%%%%%%%%%%%%%%%%%%%%%%%%%%%%
\subsection{Rectangular group $\Cvv$} \label{sec:C2v}

\begin{figure}
  \begin{minipage}{0.49\textwidth}
    \centering
    \textsf{Examples for group $\Cvv$}

    \vspace{3mm}
    \centering
    \includegraphics[width=0.85\columnwidth]{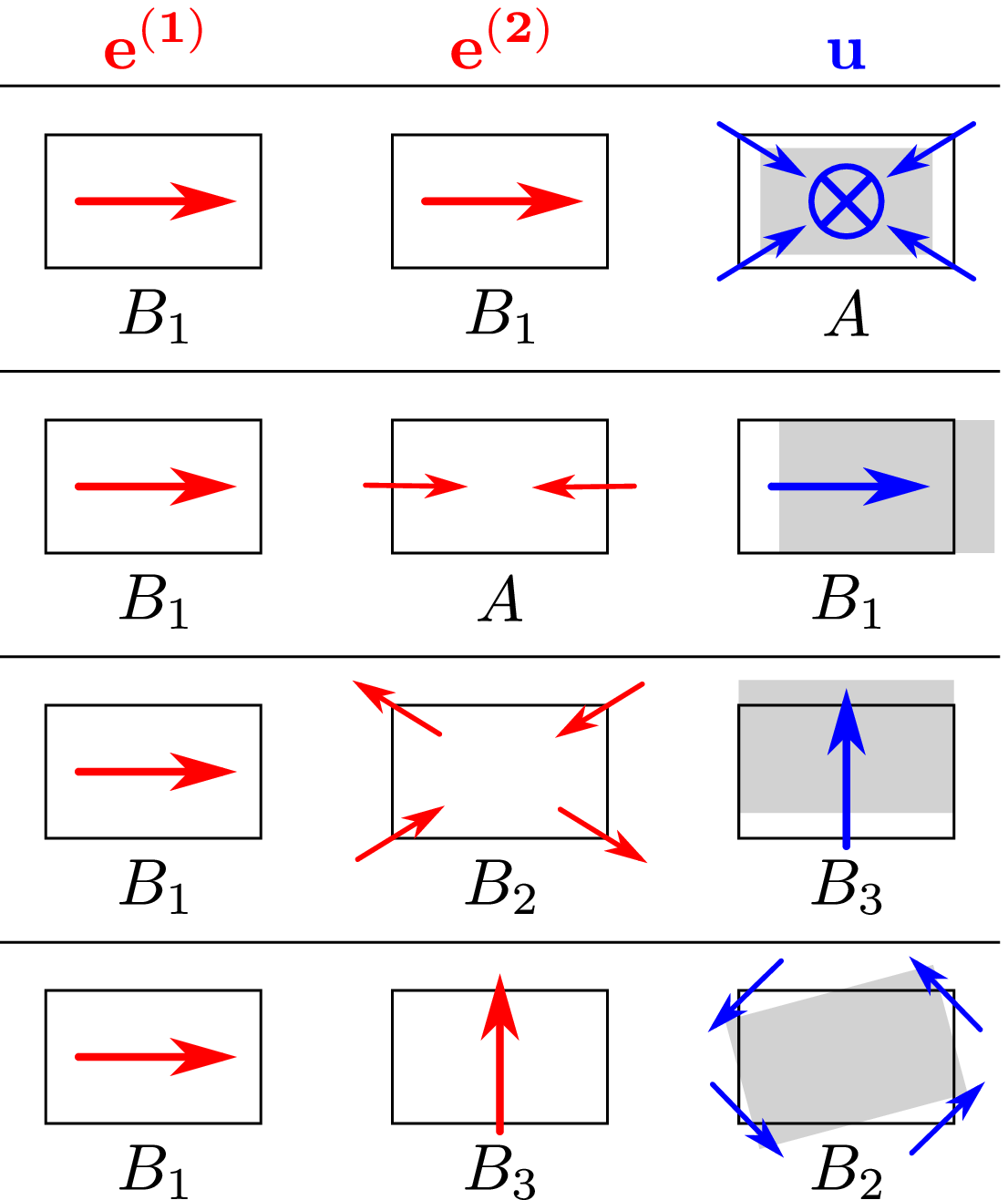}
  \end{minipage}
  \begin{minipage}{0.49\textwidth}
    \centering
    \textsf{Examples for group $\Cvvv$}

    \vspace{3mm}
    \centering
    \includegraphics[width=0.85\columnwidth]{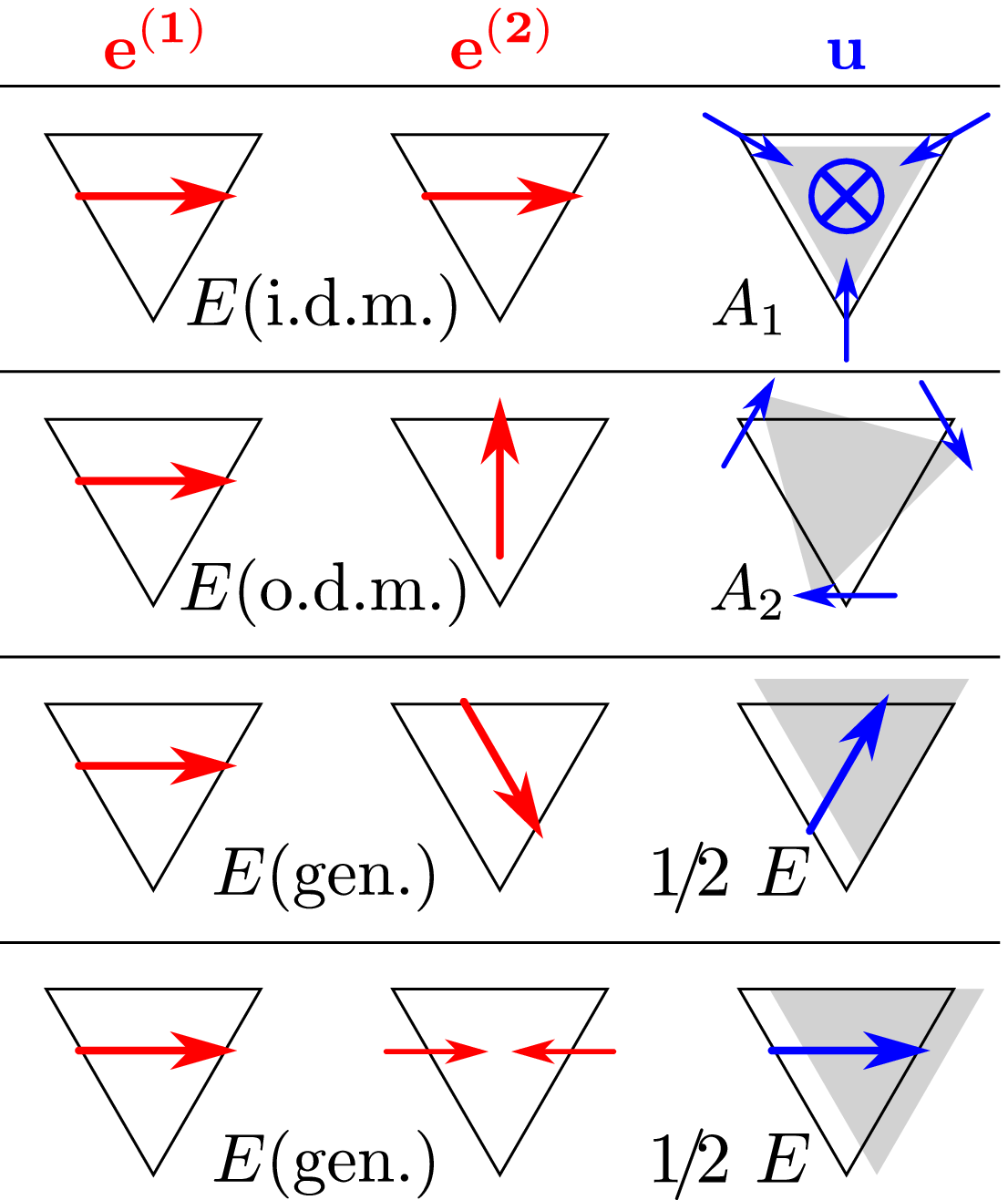}
  \end{minipage}
  \caption{
    Typical examples for combinations of symmetric eigenmodes that can have
    non-zero opto-acoustic overlap for the point groups $\Cvv$ (left hand
    side) and $\Cvvv$ (right hand side).
    Each line contains sketches of two optical modes (left and middle column)
    and one acoustic mode (right column).
    The symmetry of the undistorted waveguide is represented by a black polygon,
    the effect of distortion is sketched in light gray.
    Red arrows indicate the general behavior of the major component of the 
    optical modes' electric fields.
    Blue arrows indicate the general behavior of the acoustic displacement field.
  }
  \label{fig:C2v_and_C3v}
\end{figure}

The rectangular group does not exhibit any degenerate representations and is
therefore simple to discuss.
The symmetry of the integrand in \eqnref{eqn:integrand} can be easily established 
by multiplying the characters of the constituent representations.
We find two classes of combinations that lead to $w$ having trivial symmetry.
The first class is if two modes have identical symmetry and the third one has
$A$-symmetry.
This is the case in backward SBS, where the two optical modes have identical
symmetry (but opposite propagation direction) and is only possible with an
acoustic mode of $A$-type, \ie not with flexural or torsional modes.
Another example for this class is the conversion between a fundamental, linearly
polarized optical mode ($B_1$-type or $B_3$-type) and a higher order mode
($A$-type).
In this case, the acoustic mode must be of flexural type.
The second class is if the optical and acoustic modes cover all three 
non-trivial representations.
This is for example the case for conversion between orthogonally polarized 
fundamental optical modes ($B_1$-type and $B_3$-type) and is only possible
with an acoustic mode of $B_2$-type, \ie a torsional mode.
These combinations are sketched in the left hand part of \figref{fig:C2v_and_C3v}.

%%%%%%%%%%%%%%%%%%%%%%%%%%%%%%%%%%%%%%%%%%%%%%%%%%%%%%%%%%%%%%%%%%%%%%%%%%%%%%%%
\subsection{Triangular group $\Cvvv$} \label{sec:C3v}

The triangular point group supports one degenerate and two non-degenerate 
representations.
The degenerate representations describe linearly polarized optical and
flexural acoustic modes.
This degeneracy slightly complicates the problem of identifying gain 
combinations.
The reason for this is that the symmetry of the product of degenerate modes 
depends on the degree of overlap between them.
To resolve this, we have to treat the cases of orthogonal eigenmodes (orthogonal
polarizations) separately from the case where the two modes have maximally 
possible overlap (identical polarization).
The product decomposition table for the group $\Cvvv$ in 
\tabref{tab:prod_decomp} shows that the former case always results in a 
mode of $A_2$-symmetry, whereas the latter case leads to a superposition of 
a trivial and a degenerate representation.
The general case of partially overlapping eigenmodes (skew polarizations)
is a superposition of these special cases.
The same holds if both relevant eigenmodes stem from different degenerate
subspaces (\eg an optical and an acoustic mode of $E$-symmetry).

As before, we find two classes of gain-supporting combinations.
The first case is again the combination of one $A_1$-type eigenmodes and 
two other modes of identical symmetry properties, \ie either two modes of the
same non-degenerate type or a pair of degenerate modes with the same
polarization.
Again, this implies that backward-SBS requires an $A_1$-type, \ie a 
longitudinal acoustic mode.
The second case is again a combination of all three non-trivial symmetry
properties, \ie one $A_2$-type mode and two orthogonal modes of $E$-type.
As in the rectangular case, this means that SBS between orthogonally polarized 
optical modes requires a torsional mode.
The product between a fundamental optical mode and a flexural acoustic mode
is the general case of the product decomposition and may contain any 
representation.
This means that a flexural acoustic mode can cause SBS between a fundamental 
mode and any other optical band, not only between the fundamental and the
corresponding higher-order mode.
We provide some examples for combinations of modes with non-zero overlap
in the right hand part of \figref{fig:C2v_and_C3v}.

%%%%%%%%%%%%%%%%%%%%%%%%%%%%%%%%%%%%%%%%%%%%%%%%%%%%%%%%%%%%%%%%%%%%%%%%%%%%%%%%
\subsection{Hexagonal group $\Csixv$} \label{sec:C6v}

The hexagonal group $\Csixv$ is a supergroup of both $\Cvv$ and $\Cvvv$, and 
supports two degenerate and four non-degenerate representations. 
They emerge from the representations of the $\Cvvv$ group by including the 
$C_2$ operation as a new generator. 
The $E_1$ representations include  linear polarized (dipole-like) 
modes, which are the fundamental modes in the optical case. 
The modes with $E_2$ symmetry have quadrupolar character, as can be seen 
in the second-order optical mode provided this is above cut-off 
(see \figref{fig:optical_examples}).
The $B_1$ and $B_2$ modes have hexapolar character---in contrast to the dipolar 
and quadrupolar modes these modes 
are not degenerate---and the 
$A_2$ representation correspond to modes with dodecapolar symmetry such as
torsional modes in the acoustic case and azimuthally polarized optical modes.
As in the rectangular and triangular cases, the $A_1$ representation 
corresponds to longitudinal or breathing acoustic modes and radially polarized 
optical modes. 

The analysis of possible gain combinations is greatly simplified by the prior 
discussion of the two subgroups $\Cvv$ and $\Cvvv$. 
First we note that the character table for the non-degenerate representation 
is very similar to that of $\Cvv$.
The only difference are the rotation operators $C_3$ and $C_6$, whose 
characters are identical to those of $E$ and $C_2$ in the non-degenerate case.
As a consequence, the discussion of \secref{sec:C2v} applies to the gain 
combinations of any non-degenerate representations. 
Furthermore we notice that the representation table reduces to two separate 
instances of the $\Cvvv$ table, which differ only by the sign of the character 
of the $C_2$ rotation. 
As a consequence the discussion of \secref{sec:C3v} applies, with the 
additional constraint that the number of representations that contain a 
negative character for $C_2$ must be even in order to obtain the trivial 
representation. 

\begin{figure}
  \centering
  \textsf{Examples for group $\Csixv$}

  \vspace{3mm}
  \centering
  \includegraphics[width=0.92\columnwidth]{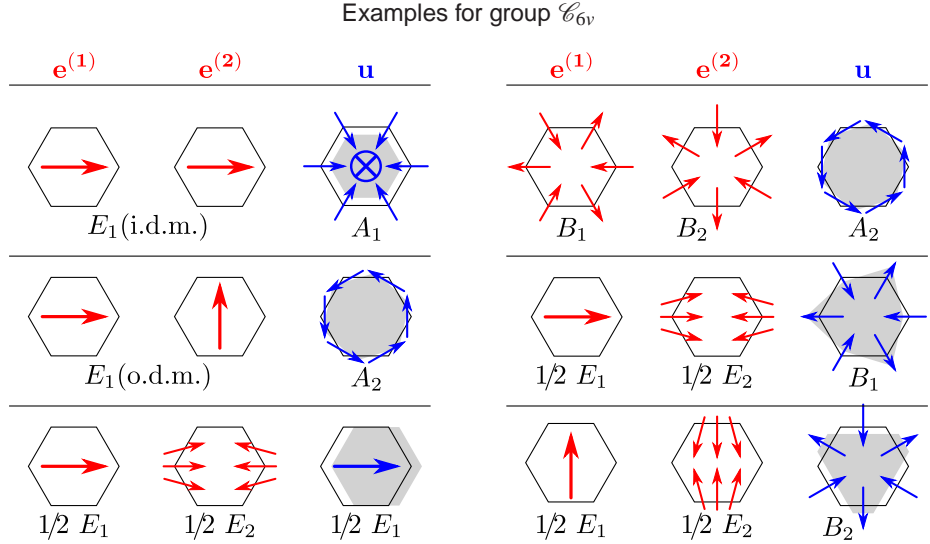}
  \caption{
    Typical examples for combinations of symmetric eigenmodes that can have
    non-zero opto-acoustic overlap for the point group $\Csixv$.
    See the caption of \figref{fig:C2v_and_C3v} for a more detailed 
    description of what is shown here.
  }
  \label{fig:C6v}
\end{figure}

The final case is unique to hexagonal structures, and consists of combinations 
of $E_1$ together with $E_2$ modes. 
This type of interaction must be handled with the help of the product 
decomposition table (see \tabref{tab:prod_decomp}). 
According to this table, two such modes can be combined with either a $B$-type 
non-degenerate mode or with an $E_1$-type mode in order to obtain non-zero gain. 
As an example, an optical fundamental mode can couple to the corresponding 
second order mode via an acoustic flexural mode. 
Alternatively, an acoustic accordion-like acoustic mode can facilitate the 
coupling between two optical modes of the same type.
A complete discussion of all the possible mode combinations would be excessive 
without being particularly enlighting; a representative selection of examples 
that might be important is given in \figref{fig:C6v}. 

%%%%%%%%%%%%%%%%%%%%%%%%%%%%%%%%%%%%%%%%%%%%%%%%%%%%%%%%%%%%%%%%%%%%%%%%%%%%%%%%
%%%%%%%%%%%%%%%%%%%%%%%%%%%%%%%%%%%%%%%%%%%%%%%%%%%%%%%%%%%%%%%%%%%%%%%%%%%%%%%%
\section{Conclusion}

We have formulated the selection rules for SBS for three important waveguide 
symmetries. 
If the symmetry is only weakly broken, for example by anisotropy in the 
mechanical properties induced by the cubic nature of some materials, we can 
expect these selection rules to hold to a good approximation. 
If the symmetry is strongly broken, these rules are readily adapted to 
waveguides with reduced symmetry---for example, the selection rules for
rectangular waveguides on a substrate will reduce to the $\Cv$ group, 
combinations of which will follow the character table with the $\sigma_y$ and 
$C_2$ operations removed.

Apart from its use in the understanding and design of SBS-active waveguides,
our results are very useful for the development and testing of numerical tools.
Firstly, they provide sets of modes that should not be able to couple for
symmetry reasons.
This is an excellent way to identify convergence and discretization issues in
the evaluation of the coupling integrals.
Secondly, they eliminate complete symmetry classes of acoustic solutions for the
interaction between any two given optical eigenmodes.
These symmetry constraints can be exploited in numerical calculations by 
reducing the computational domain and imposing appropriate boundary conditions
along the mirror planes, thereby greatly reducing the computational effort.
This can be significant in automatized design optimazation problems or 
large-scale geometries.

%%%%%%%%%%%%%%%%%%%%%%%%%%%%%%%%%%%%%%%%%%%%%%%%%%%%%%%%%%%%%%%%%%%%%%%%%%%%%%%%
%%%%%%%%%%%%%%%%%%%%%%%%%%%%%%%%%%%%%%%%%%%%%%%%%%%%%%%%%%%%%%%%%%%%%%%%%%%%%%%%
\section*{Acknowledgments}

\noindent
We acknowledge financial support of the Australian Research Council via
Discovery Grant DP130100832. 

\end{document}